# A hierarchical Bayesian approach for estimating the origin of a mixed population[*]

## Feng Guo[1], Dipak K. Dey[2] and Kent E. Holsinger[3]

*Virginia Polytechnic Institute, University of Connecticut and University of Connecticut*

**Abstract:** We propose a hierarchical Bayesian model to estimate the proportional contribution of source populations to a newly founded colony. Samples are derived from the first generation offspring in the colony, but mating may occur preferentially among migrants from the same source population. Genotypes of the newly founded colony and source populations are used to estimate the mixture proportions, and the mixture proportions are related to environmental and demographic factors that might affect the colonizing process. We estimate an assortative mating coefficient, mixture proportions, and regression relationships between environmental factors and the mixture proportions in a single hierarchical model. The first-stage likelihood for genotypes in the newly founded colony is a mixture multinomial distribution reflecting the colonizing process. The environmental and demographic data are incorporated into the model through a hierarchical prior structure. A simulation study is conducted to investigate the performance of the model by using different levels of population divergence and number of genetic markers included in the analysis. We use Markov chain Monte Carlo (MCMC) simulation to conduct inference for the posterior distributions of model parameters. We apply the model to a data set derived from grey seals in the Orkney Islands, Scotland. We compare our model with a similar model previously used to analyze these data. The results from both the simulation and application to real data indicate that our model provides better estimates for the covariate effects.

## Contents

---

[*]Supported by NIH Grant 1R01-GM068449-01A1.

[1]Department of Statistics, Virginia Polytechnic Institute, Blacksburg, VA 24060, USA, e-mail: feng.guo@vt.edu
[2]Department of Statistics, University of Connecticut, Storrs, CT 06269, USA, e-mail: dey@stat.uconn.edu
[3]Department of Ecology and Evolutionary Biology, University of Connecticut, Storrs, CT 06269, USA, e-mail: kent@darwin.eeb.uconn.edu

*AMS 2000 subject classifications:* Primary 60K35, 60K35; secondary 60K35.

*Keywords and phrases:* hierarchical Bayes, MCMC, multinomial.

## 1. Introduction

Fisheries scientists and marine biologists are often faced with the problem of identifying proportions of individuals in a single catch that come from different stocks. Estimating these proportions is necessary for evaluating the effect of commercial fisheries on individual fisheries stocks and for understanding the ecological factors that influence the relative contributions of different stocks. Similarly, those who study marine mammals are often interested in identifying the source populations for newly founded colonies as well as environmental or demographic factors that influence the relative contributions of different sources. The increasing ease with which genetic data are collected and the tendency for populations of species to become genetically differentiated over time has led to the increase in using genetic markers to estimate the proportional contribution of source populations to mixed stocks. The rationale is simple: allele frequencies are likely to differ among source populations, and genotype frequencies in the harvest site/new habitat are determined by the proportional contributions of the source populations. Both the differences among source populations and the mixture proportions can be detected by appropriate statistical models.

Several methods have been developed for the inference of the proportional contribution, $\mathbf{m}$, where $m_i$ is the percentage of individuals in the mixed population originating from source $i$. Conditional Maximum Likelihood Estimates (MLEs) have been widely used [8, 9]. The conditional MLE assumes the sampled source populations are exhaustive lists of all possible sources and the allele frequency of the sources are known without error. Neither assumption is satisfied for real samples. Smouse et al. [12] extended conditional MLEs to unconditional MLEs in which the source allele frequencies are treated as estimated parameters and unknown sources are allowed to be presented in the model.

In recent years, several authors have applied Bayesian methods to the stock mixture problem [2, 3, 4, 9, 11]. One advantage of a Bayesian model is that the influence of the genotype data from the mixture population on the estimation of source allele frequency is fully incorporated through the joint likelihood and is reflected in the posterior distribution of allele frequencies. Another advantage is that we can include non-genetic information in the model through appropriate priors. We can, for example, set the prior distribution of mixture proportions as a function of ecological or demographic parameters.

Until recently interest has largely focused on inference of the proportional stock contributions, but there is increasing interest in understanding the ecological factors that influence those proportions, e.g., source population size and the distance between the source and mixture habitat [3, 9]. While including these relationships is difficult to implement in classical models, a hierarchical Bayesian model can easily incorporate these relationships into the prior for the vector of proportional contributions, $\mathbf{m}$. Existing approaches for inference on $\mathbf{m}$ either use an additive logistic transformation with parameters assumed to have normal priors on the logistic scale [9] or model $\mathbf{m}$ directly on the simplex using a Dirichlet with parameters assumed to be lognormal [3]. In this paper, we propose an alternative formulation of Dirichlet prior structure that is both more efficient in separating mean and variance effects and also directly interpretable. We conduct a simulation study to demonstrate our approach and investigate its performance by varying the level of differentiation among source populations and number of genetic markers. We apply our model to the data derived from grey seals in the Orkney Islands and compare our results with those obtained with a Dirichlet–lognormal prior [3] and with a model using a uniform prior for $\mathbf{m}$.

## 2. Models

We conduct our analysis in a Bayesian framework. The parameters, such as (relative) allele frequency, $\mathbf{P}$, and proportional contribution, $\mathbf{m}$, are considered as random variables and the statistical inference is based on the posterior distributions of parameters. In this analysis, the genetic data from source and mixture populations are included in the likelihood function and the covariate information is included through a hierarchical prior structure.

The likelihood of the data is derived from genetic theory. The genetic data consist of two parts: the allele counts from source populations and genotype counts from the mixed population. Gaggiotti's model [3] deals with the situation where there is one new colony and several source populations that might contribute to the founding group of the new colony. The data are collected from the first generation descendants of migrants, but the model allows for non-random, assortative mating, i.e., individuals from the same source population are more likely to mate with one another than those from different source populations.

Consider a first generation descent individual $k$ in the new colony whose mother is from population $i$ and father is from population $j$. Denote $P(\mathbf{y}_k|ij)$ as the probability that this individual has genotype $\mathbf{y}_k$, which includes $L$ loci. Denote $(a_{1lk}, a_{2lk})$ as the genotype of individual $k$ at locus $l$. First consider individuals with both parents from the same population, i.e., $i = j$. Assume mating is random among those individuals and Hardy–Weinberg Equilibrium (HWE), which states that the frequency of the heterozygous genotype is twice that of the homozygous genotype, holds. The probability of genotype $\mathbf{y}_k$ is,

$$(2.1) \qquad P(\mathbf{y}_k|ii) = \prod_{l=1}^{L} \delta_{lk} p_{a_{1lk;li}} p_{a_{2lk;li}},$$

where $p_{a_{1lk;li}}$ is the allele frequency of $a_{1lk}$ at locus $l$ in population $i$, and $\delta_{lk}$ is an indicator variable defined as

$$\delta_{lk} = \begin{cases} 1, & \text{if } a_{1lk} = a_{2lk}, \\ 2, & \text{if } a_{1lk} \neq a_{2lk}. \end{cases}$$

When the parents are from different populations, i.e., $i \neq j$, $P(\mathbf{y}_k|ij)$ is given by

$$(2.2) \qquad P(\mathbf{y}_k|ij) = \prod_{l=1}^{L} (p_{a_{1lk;li}} p_{a_{2lk;lj}} + \gamma_{lk} p_{a_{2lk;li}} p_{a_{1lk;lj}}),$$

where

$$\gamma_{lk} = \begin{cases} 0, & \text{if } a_{1lk} = a_{2lk}, \\ 1, & \text{if } a_{1lk} \neq a_{2lk}. \end{cases}$$

Parameter $\gamma_{lk}$ indicates that when alleles at a locus are different, there are two different ways of assigning them to parents in different source populations.

When mating happens assortatively, i.e., individuals tend to mate with those from the same source population, HWE is not valid. The assortative mating can be modeled by an assortative mating coefficient $\omega \in (0,1)$. Specifically, a proportion $\omega$ of first generation descendants arise from assortative mating among individuals from the same source and a proportion $1 - \omega$ arise from random mating among all migrants. Consequently, the likelihood of finding the genotype $\mathbf{y}_k$ in a sample from the new colony is as follows:

$$
\begin{aligned}
P(\mathbf{y}_k|\omega, \mathbf{P}, \mathbf{m}) = &\omega \sum_{i=1}^{I} m_i P(\mathbf{y}_k|ii) + \\
(2.3) \qquad &(1 - \omega) \left[ \sum_{i=1}^{I} m_i^2 P(\mathbf{y}_k|ii) + \sum_{i=1}^{I} \sum_{j \neq i} m_i m_j P(\mathbf{y}_k|ij) \right],
\end{aligned}
$$

where $P(\mathbf{y}_k|ii)$ and $P(\mathbf{y}_k|ij)$ is as in (2.1) and (2.2).

Since HWE is assumed for source populations, the genotype frequency is determined by the allele frequency. It is easy to show that the likelihood associated with the genotype frequencies is equivalent to a multinomial with parameters corresponding to the allele frequencies and response variables as allele counts from source populations. If we assume independence among the genotype counts across loci and populations, the likelihood function for source allele counts is a product multinomial:

$$
P(\mathbf{N}|\mathbf{P}) \sim \prod_{i=1}^{I} \prod_{l=1}^{L} \prod_{j=1}^{A_l} p_{jli}^{N_{jli}},
$$

where $N_{jli}$ is the allele count for source population $i$ at locus $l$ for allele $a_j$ and $A_l$ is the number of alleles at locus $l$.

The prior distributions for $\omega$ and $\mathbf{P}$ reflect prior beliefs about plausible values for these parameters. We choose vague/noninformative priors to reflect the fact that we have no reason to expect particular values for these parameters. In particular, we assume an uniform prior on $(0, 1)$ for the assortative mating coefficient $\omega$. For allele frequencies in the source populations, $\mathbf{P}$, we assume a Dirichlet prior

$$
\pi(\mathbf{p}|\alpha) \propto \prod_{i=1}^{I} \prod_{l=1}^{L} \prod_{j=1}^{A_l} p_{jli}^{\alpha_{jl}-1}.
$$

As there is no previous data or preference for $\mathbf{P}$, it is reasonable to take $\alpha$'s all equal to 1, leading to a symmetric Dirichlet prior with parameter 1.

The key to this analysis is how to incorporate the demographic/environmental factors into the estimation of the proportional contribution $\mathbf{m}$. Information on $\mathbf{m}$ is obtained only indirectly through its influence on genotype frequencies in the mixed population. Thus, there is no simple data likelihood connecting $\mathbf{m}$ and the covariates. In a Bayesian framework, however, we can assign an informative prior for $\mathbf{m}$ containing the information from the covariates. The challenge in doing a regression type analysis is that the sum of the components of $\mathbf{m}$ must be equal to 1. Since covariates have to be considered for every source, an ordinary linear model or logit transformation does not fit here. Okuyama and Bolker [9] overcome this problem by using an additive log ratio transformation based on an additive logistic normal distribution. However, a baseline population has to be selected and the covariates need to be adjusted according to the baseline population, which makes it hard to accommodate multiple covariates and the interpretation of the coefficients is not straightforward. Gaggiotti et al. [3] use a hierarchical Dirichlet prior to address this problem. In Gaggiotti's setup, the first level prior distribution for $\mathbf{m}$ is a Dirichlet distribution in which the individual parameters follow a lognormal distribution, i.e.,

$$\mathbf{m} \sim \mathcal{D}(\boldsymbol{\psi}),$$

$$(2.4) \qquad \log(\psi_i) \sim \mathcal{N}(\mu_i, \sigma^2),$$

$$\mu_i = \alpha_0 + \sum_{r=1}^{p} \alpha_r G_{ri},$$

where $G_{ri}$ is the value of the $r$th factor for source population $i$ and $\boldsymbol{\alpha} = (\alpha_0, \ldots, \alpha_p)$ is the vector of regression coefficients. The value of the covariates are standardized and the prior for $\alpha_r$ is $\mathcal{N}(0, \sigma_p^2)$. The covariates affect the prior through the parameter $\boldsymbol{\psi}$.

We introduce a new hierarchical prior structure for the mixture proportions $\mathbf{m}$. The first level prior for $\mathbf{m}$ is a Dirichlet distribution with parameters $((1 - \rho)/\rho)\boldsymbol{\varphi}$, where $\rho \in (0, 1)$ and $\boldsymbol{\varphi}$ are the hyperparameters of the prior subject to the constraint $\sum_{i=1}^{I} \varphi_i = 1$. This form of prior is widely used in population genetics for its relationship with the measure of population differentiation, e.g., Wright's $F_{ST}$ [1, 5, 6]. Due to the fact that the covariates are also observed, another hierarchical level is added to incorporate the randomness. A Dirichlet prior with parameter $\boldsymbol{\eta}$ is assigned to $\boldsymbol{\varphi}$. The covariates are included in the model by setting the logarithm of $\boldsymbol{\eta}$ to be a function of a linear combination of the covariates, i.e.,

$$(2.5) \qquad \begin{aligned} \mathbf{m} &\sim \mathcal{D}(\frac{1 - \rho}{\rho}\boldsymbol{\varphi}), \\ \boldsymbol{\varphi} &\sim \mathcal{D}(\boldsymbol{\eta}), \\ \log(\eta_i) &= \alpha_0 + \sum_{r=1}^{p} \alpha_r G_{ri}. \end{aligned}$$

Since the covariates are normalized, the regression coefficients $\alpha_r$'s are assumed to be independent of each other. Normal priors with mean zero and a large variance, $\sigma_p^2 = 10$, are assigned to parameter $\boldsymbol{\alpha}$. The full model is:

$$(2.6) \qquad \begin{aligned} &\pi(\mathbf{P}, \omega, \mathbf{m}, \rho, \boldsymbol{\varphi}, \boldsymbol{\alpha} | \mathbf{Y}, \mathbf{N}) \\ &\propto p(\mathbf{Y} | \mathbf{P}, \omega, \mathbf{m}) \pi(\mathbf{m} | \rho, \boldsymbol{\varphi}) \pi(\boldsymbol{\varphi} | \boldsymbol{\alpha}) \pi(\boldsymbol{\alpha}) \pi(\omega) p(\mathbf{N} | \mathbf{P}) \pi(\mathbf{P}), \end{aligned}$$

where $\mathbf{Y}$ is the genotype data of the new colony, $\mathbf{N}$ is the allele count in source populations, and $\mathbf{P}$ is the allele frequency. The Directed Acyclic Graphs (DAGs) of the Dirichlet–Dirichlet and Dirichlet–Lognormal models are presented in Figure 1 and Figure 2.

The prior structure of our model puts the support of $\rho$ between $(0, 1)$ and the value of $((1 - \rho)/\rho)\varphi_i$ on the entire positive line. This setup brings several advantages. First, a natural vague prior for $\rho$ is simply a uniform distribution between 0 and 1, $\mathcal{U}(0, 1)$. Second, when we use $\mathcal{U}(0, 1)$ as a prior, the posterior mean of $\rho$ can be used as an indicator of the dispersion of the regression of $\mathbf{m}$ on demographic/environmental factors. The variance associated with component $m_i$ is $\varphi_i(1 - \varphi_i)\rho$, and $1 - \rho$ is roughly the proportion of variance in $\mathbf{m}$ explained by the regression (cf. [6]). Third, an informative prior on $\rho$ can be used to influence the variance of the prior and the relative weight of environmental covariates and genetic data on the posterior of **m**. To see this, observe that $(1 - \rho)/\rho$ increases when $\rho$ decreases and the variance of $\mathcal{D}((1 - \rho)/\rho)\boldsymbol{\varphi})$ is decided by the absolute value of $((1 - \rho)/\rho)\boldsymbol{\varphi}$. Thus a small value of $\rho$ corresponds to a small variance. As the variance of the prior usually decides the relative weight of the prior information on the posterior distribution, $\rho$ indicates the relative weight of the covariates in the posterior distribution of **m**.

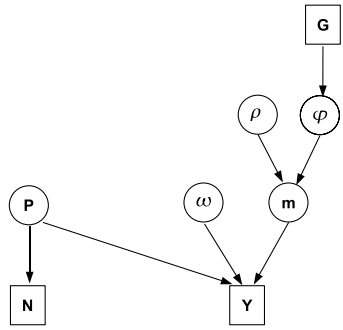

Fig 1. *Dirichlet–Dirichlet model.*

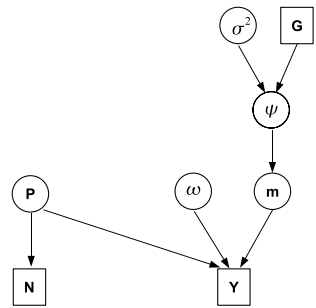

Fig 2. *Dirichlet–Lognormal model.*

In the models of [3], [9], and the model proposed in this paper, the effect of the covariates is incorporated through the prior for **m**. The parameter of interest, **m**, is determined by both genetic information through the likelihood and demographic/environmental information through the prior. An important question is, what is the relative influence of these two sources of information on posterior inference? The influence of the prior is usually directly related to its variance: with a large variance the posterior is dominated by the likelihood while with a small variance the posterior is dominated by the prior. For the Dirichlet–lognormal prior in (2.4), the mean of **m** is controlled by the value of $\boldsymbol{\psi}$. At the same time the variance of **m** is affected by both the magnitude of $\boldsymbol{\psi}$ and the distribution of $\sigma^2$. The interaction among these two parameters tends to increase the uncertainty in the posterior distribution. In contrast, the prior proposed in (2.5) clearly separates the roles of the parameters: the mean of **m** is determined by $\boldsymbol{\varphi}$ and the variance of **m** is controlled by $\rho$. This separation is due to the constraint that $\boldsymbol{\varphi}$ is on the simplex. In the simulation study and the application to a real life dataset, we illustrate that the Dirichlet–Dirichlet prior shows less variation in estimating the covariate effects while providing comparable coverage of interval estimates.

## 3. Simulation study

We conduct a simulation study to investigate the performance of our proposed model by simulating data from populations with different levels of genetic differentiation as well as different numbers of genetic markers. As discussed above, the estimation of the mixture proportions relies on the divergence among source populations. We are interested in how different level of divergence among source populations would affect the posterior distributions of the parameters of interest. From a practical point of view, population divergence level cannot be controlled by researchers. Instead, researchers can determine how many genetic markers are to be assigned and included in the analysis. Hence, we are also interested in the relationship between the number of loci and the posterior distribution of the parameters.

We consider three simulation scenarios. Under the first scenario, the level of population differentiation is moderate and there are a relatively small number of genetic markers, e.g., 8 loci are available. Under the second scenario, the number of loci is the same as the first scenario but the level of genetic differentiation among source populations is higher. Under the last scenario, the genetic variation is the same as that of the first scenario but the number of genetic markers is doubled, i.e., 16 genetic markers are available. The number of source populations, number of individuals in the mixed population, and allele counts in the source populations are comparable with those in the grey seal data we analyze later.

In the first part of the simulation we generate allele counts in the source populations, which should reflect the level of genetic differentiation among them. This is realized through a hierarchical population structure. We assume that the allele frequencies of the source populations are from a common hyper-population, which has fixed allele frequencies $\boldsymbol{\psi}$, a $L \times A$ matrix with $L$ being the number of loci and $A$ being the number of alleles at each locus. (Without loss of generality, we assume all loci have the same number of alleles.) The allele frequencies $\mathbf{p}_{li}$, a $1 \times A$ vector, for source $i$ and locus $l$ are random samples from a Dirichlet distribution, $\mathcal{D}(((1-\theta)/\theta)\boldsymbol{\psi}_l)$, where $\boldsymbol{\psi}_l$, a $1 \times A$ vector, is the allele frequency of locus $l$ for the hyper-population, and $\theta$ is a population divergence measure used widely in population genetic studies, namely Wright's $Fst$. Note that $E[p_{jli}] = \psi_{jl}$ and $Var[p_{jli}] = \theta \psi_{jl}(1 - \psi_{jl})$, where $\psi_{jl}$ is the allele frequency of locus $l$, allele $j$ for the hyper-population. We choose $\theta = 0.05$ and $\theta = 0.2$ for small and large divergence scenarios, respectively.

The detailed simulation is described as follows. Step 1: generate allele frequencies of the hyper population, $\boldsymbol{\psi}$, by generating a random sample $L$ times from an $A$ dimensional symmetric Dirichlet distribution with parameter 1. Step 2: generate allele frequencies, $\mathbf{p}_{li}$, from the Dirichlet distribution $\mathcal{D}(((1-\theta)/\theta)\boldsymbol{\psi}_l)$ with predefined $\theta$. Step 3: generate allele counts, $N_{li}$, for source $i$ and locus $l$, from a multinomial distribution with total allele counts $N = 400$, and probability $\mathbf{p}_{li}$ (from step 2).

In the second part of the simulation we generate genotypes of individuals from the mixed population, which requires the proportional contributions $\mathbf{m}$ and the probability of each genotype. We adopt fixed proportional contributions, which are a function of the two covariates. Note that in both the Dirichlet–Dirichlet (2.5) and the Dirichlet–lognormal (2.4) models, the conditional expectations of the prior for $\mathbf{m}$ are the same, namely,

$$E[m_i|\boldsymbol{\alpha}] = \frac{e^{\boldsymbol{\alpha} \cdot \mathbf{G}_i}}{\sum_{i=1}^{I} e^{\boldsymbol{\alpha} \cdot \mathbf{G}_i}},$$

where $\boldsymbol{\alpha}$ is the vector of regression coefficients and $\mathbf{G}_i$ is the vector of covariates for source $i$. We use two covariates with the values shown in Table 1 and the coefficients are set to $\alpha_1 = -0.5$ and $\alpha_2 = 0.5$.

Table 1
*Normalized covariates*

| Source | Distance | Productivity |
|--------|----------|--------------|
| 1 | $-0.295$ | 1.298 |
| 2 | $-0.849$ | 1.285 |
| 3 | $-0.822$ | $-0.238$ |
| 4 | $-0.562$ | $-1.256$ |
| 5 | $-0.326$ | $-0.729$ |
| 6 | 1.533 | 0.286 |
| 7 | 1.320 | $-0.646$ |

The genotype of an individual $k$ is generated by the following steps. First, we decide whether its parents are from the same source by comparing a uniform random number on [0,1] with a preset assortative coefficient $w = 0.05$. The second step is to generate the genotype frequency at each locus. If the parents are from the same source population, the probability of genotype $\mathbf{y}_k$ is $\sum_{i=1}^{I} m_i P(\mathbf{y}_k|ii)$, where $P(\mathbf{y}_k|ii)$ is as in (2.1). If the parents come randomly from the source populations then the probability of genotype $\mathbf{y}_k$ is $\sum_{i=1}^{I} m_i^2 P(\mathbf{y}_k|ii) + \sum_{i=1}^{I} \sum_{j \neq i} m_i m_j P(\mathbf{y}_k|ij)$, where $P(\mathbf{y}_k|ij)$ is the probability of parents from different source populations as in (2.2). Once we have the probability of each genotype for individual $k$ at locus $l$, we can easily generate the genotype from this probability. Step 2 is repeated for each locus of the individual to get the complete genotype of individual $k$. The above steps are repeated 160 times to get the genotypes of 160 individuals in the mixed population.

We generate 50 data sets for each of the three scenarios, and we fit both the Dirichlet–Dirichlet prior (2.5) proposed in this paper and the Dirichlet–lognormal prior (2.4) to each data set using a MCMC method. Since most of the parameters are vectors on a simplex, we use a multi-dimensional logit transformation to put the support of the transformed parameters on the real line and remove the simplex constraint. A normal proposal density is then used to conduct a Metropolis–Hastings update nested in the Gibbs sampling. Details of the MCMC update procedure are presented in the Appendix. For most of the data set, we conduct 30,000 iterations in the simulation with 5,000 burn-in and thin the MCMC output by 5. For chains showing suspicious convergence behavior, longer iterations and fine tuning are used to ensure convergence.

Table 2 presents a summary of the posterior analysis, including the average of the posterior means, posterior standard deviations, root mean square error (RMSEs), and the lengths of the 95% highest probability density (HPD) intervals. In general, the posterior means of $\mathbf{m}$ are reasonably close to the true values in all scenarios. The effects of population divergence and number of loci are reflected mainly in the posterior dispersion of $\mathbf{m}$. As shown in the Table, the lengths of the 95% HPD intervals, the posterior standard deviations, and the RMSEs, all indicate that the posterior dispersion of $\mathbf{m}$ decreases with the increase of population differentiation. Given the same level of population differentiation, increasing the number of genetic markers also significantly improves the precision of posterior estimation for $\mathbf{m}$. These results suggest that although in practice the population divergence is always fixed, collecting and including more genetic markers in the analysis can significantly improve the estimation of the proportional contribution parameters $\mathbf{m}$.

For the regression coefficient $\boldsymbol{\alpha}$, both models provide reasonable estimates for the posterior means. However, the posterior variation is large and the 95% HPD

Table 2
*Posterior summary of simulation study*

| | TRUE | Dirichlet–Dirichlet | | | Dirichlet–lognormal | | |
|---|---|---|---|---|---|---|---|
| | | $\theta = 0.05$ L=8 | $\theta = 0.20$ L=8 | $\theta = 0.05$ L=16 | $\theta = 0.05$ L=8 | $\theta = 0.2$ L=8 | $\theta = 0.05$ L=16 |
| $m_1$ | 0.249 | *0.256 | 0.254 | 0.239 | 0.258 | 0.255 | 0.239 |
| | | **0.044 | 0.033 | 0.036 | 0.045 | 0.032 | 0.036 |
| | | ***0.063 | 0.052 | 0.053 | 0.065 | 0.052 | 0.054 |
| | | ****0.170 | 0.127 | 0.138 | 0.173 | 0.126 | 0.139 |
| $m_2$ | 0.327 | 0.326 | 0.377 | 0.350 | 0.328 | 0.379 | 0.351 |
| | | 0.045 | 0.036 | 0.039 | 0.046 | 0.036 | 0.039 |
| | | 0.068 | 0.076 | 0.058 | 0.069 | 0.079 | 0.059 |
| | | 0.175 | 0.140 | 0.150 | 0.178 | 0.141 | 0.151 |
| $m_3$ | 0.151 | 0.134 | 0.125 | 0.137 | 0.135 | 0.125 | 0.138 |
| | | 0.038 | 0.026 | 0.031 | 0.038 | 0.026 | 0.031 |
| | | 0.052 | 0.053 | 0.048 | 0.052 | 0.053 | 0.049 |
| | | 0.145 | 0.099 | 0.121 | 0.149 | 0.100 | 0.121 |
| $m_4$ | 0.079 | 0.073 | 0.065 | 0.077 | 0.072 | 0.064 | 0.075 |
| | | 0.033 | 0.020 | 0.027 | 0.033 | 0.020 | 0.026 |
| | | 0.042 | 0.038 | 0.039 | 0.044 | 0.038 | 0.039 |
| | | 0.119 | 0.075 | 0.103 | 0.121 | 0.074 | 0.100 |
| $m_5$ | 0.092 | 0.094 | 0.078 | 0.090 | 0.094 | 0.077 | 0.090 |
| | | 0.033 | 0.022 | 0.027 | 0.034 | 0.021 | 0.028 |
| | | 0.051 | 0.036 | 0.044 | 0.051 | 0.037 | 0.044 |
| | | 0.1259 | 0.082 | 0.103 | 0.126 | 0.082 | 0.105 |
| $m_6$ | 0.060 | 0.069 | 0.060 | 0.060 | 0.066 | 0.059 | 0.059 |
| | | 0.033 | 0.020 | 0.025 | 0.033 | 0.020 | 0.024 |
| | | 0.045 | 0.032 | 0.038 | 0.046 | 0.033 | 0.038 |
| | | 0.115 | 0.074 | 0.090 | 0.114 | 0.074 | 0.089 |
| $m_7$ | 0.042 | 0.047 | 0.041 | 0.048 | 0.047 | 0.040 | 0.048 |
| | | 0.025 | 0.015 | 0.022 | 0.026 | 0.016 | 0.0232 |
| | | 0.038 | 0.029 | 0.037 | 0.039 | 0.029 | 0.036 |
| | | 0.088 | 0.057 | 0.078 | 0.089 | 0.056 | 0.079 |
| $\alpha_1$ | −0.500 | −0.398 | −0.429 | −0.430 | −0.485 | −0.493 | −0.493 |
| | | 0.477 | 0.469 | 0.479 | 0.613 | 0.578 | 0.587 |
| | | 0.516 | 0.496 | 0.503 | 0.659 | 0.615 | 0.625 |
| | | 1.849 | 1.815 | 1.849 | 2.383 | 2.268 | 2.305 |
| $\alpha_2$ | 0.500 | 0.449 | 0.520 | 0.433 | 0.538 | 0.618 | 0.515 |
| | | 0.416 | 0.407 | 0.408 | 0.535 | 0.519 | 0.523 |
| | | 0.437 | 0.425 | 0.422 | 0.555 | 0.551 | 0.535 |
| | | 1.637 | 1.600 | 1.602 | 2.097 | 2.041 | 2.056 |
| $\omega$ | 0.050 | 0.037 | 0.014 | 0.014 | 0.037 | 0.014 | 0.014 |
| | | 0.035 | 0.013 | 0.014 | 0.035 | 0.013 | 0.014 |
| | | 0.040 | 0.039 | 0.038 | 0.040 | 0.039 | 0.038 |
| | | 0.106 | 0.040 | 0.042 | 0.106 | 0.040 | 0.042 |

*: average of posterior means;
**: average of posterior standard deviations;
***: average of RMSEs;
****: average length of 95% HPD intervals.

intervals all contain zero. Results from the simulation study indicate that neither level of population differentiation nor number of loci has significant effects on the precision of **α** estimates. We consider this as a reasonable result since the covariate coefficients are essentially a regression over 7 data points, i.e., the 7 source populations. The level of divergence and number of loci improve the precisions of the posterior variance sfor **m**, which only affect **α** indirectly. With only 7 data points, few simulated data sets will be able to provide strong support for a regression re-lationship. Thus, increasing the number of loci or studying highly differentiated populations will do little to improve posterior estimates of $\boldsymbol{\alpha}$. A larger number of populations would be required to provide statistically supportable evidence of the effects.

The advantages of the Dirichlet–Dirichlet prior proposed in this paper are seen primarily in the reduced posterior variation of the regression coefficients, $\alpha_1$ and $\alpha_2$. Under the Dirichlet–Dirichlet prior, the posterior standard deviations and RMSEs are uniformly smaller than that of the Dirichlet–lognormal model even though the prior variances for $\boldsymbol{\alpha}$ are all set to the same value, i.e., $\sigma_p^2 = 10$. We consider this as mainly due to the confounding of the effects that both $\boldsymbol{\psi}$ and $\sigma^2$ have on the variances of $\mathbf{m}$. Another possible reason is the effects of the prior for $\tau = 1/\sigma^2$. In any case, the Dirichlet–Dirichlet prior has the advantage of leading to more precise estimation of regression coefficients and ease in picking a non-informative prior without sacrificing nominal coverage of credible intervals.

## 4. Application to the grey seal data set

To illustrate the usefulness of our approach, we apply it to data from grey seal, *Helicoerus grypus*, populations in the Orkney Islands, which were also analyzed by [3] and [4]. The data consist of allele frequencies of 8 loci for seven source colonies and the genotype frequencies for a newly established colony on Stronsay island. There are two explanatory variables associated with each source population: distance between the source island and Stronsay island ($\alpha_1$), and the 'productivity' index, which is related to the population density and size of the source population, ($\alpha_2$). The genetic data were collected from the first generation descendants of migrants to Stronsay. We use the likelihood in equation (2.3) to allow for the possibility that migrants are more likely to mate with other individuals from the island from which they migrated. Since there is no closed form for the posterior distribution, we use MCMC methods for posterior inference.

We compare results from three models with different priors: the Dirichlet–Dirichlet prior, the Dirichlet–Lognormal prior, and a model with the symmetric Dirichlet prior with parameter 1 for $\mathbf{m}$, which corresponds to a model in which covariate effects are not incorporated. Our results reveal that differences among the models rarely lead to substantial differences in the mean posterior likelihood, which is intuitively reasonable. The part of the likelihood function concerned with source population allele frequencies is identical across all models, and the part of the likelihood concerned with colony allele frequencies is tightly tied to the observed genotypes. The three models differ only through the prior for the proportional contributions $\mathbf{m}$, which has limited impact on the likelihood unless the source population differs substantially. A direct consequence of these properties is the similarity among metrics for model evaluation measures that use only the likelihood, e.g., DIC [13] and the logarithm of the pseudomarginal likelihood (LPML) [7]. As shown in Table 3, neither DIC nor LPML provides strong support for any of the models relative to the others.

The posterior densities of the model parameters are given in Figure 3. Table 4 gives the posterior means and 95% HPD intervals of the parameters of interest: $\mathbf{m}, \tau, \boldsymbol{\alpha}$ and $\rho$. It can be seen that the posterior means of $\mathbf{m}$ are quite different for the model with symmetric Dirichlet prior (with no covariates) and the models using covariate information. Specifically, models using covariate information suggest a larger proportion from sources 2 and 3 than the uniform model, which is a reasonable result since sources 2 and 3 are the closest source islands to the new colony

TABLE 3
*Model evaluation*

| Models | Dbar | pD | DIC | LPML |
|---|---|---|---|---|
| Dirichlet–Dirichlet | 8044 | 336 | 8380 | −3023 |
| Dirichlet–lognormal | 8042 | 337 | 8379 | −3023 |
| Uniform | 8045 | 336 | 8381 | −3023 |

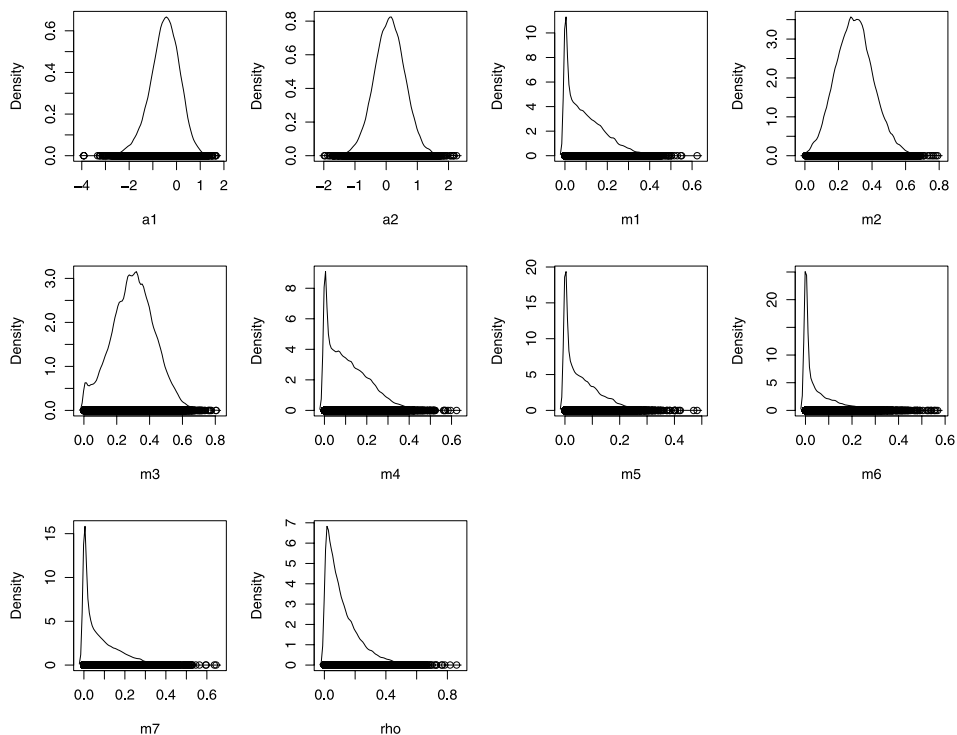

FIG 3. *The posterior densities of parameters in the Dirichlet–Dirichlet model.*

TABLE 4
*Posterior means and 95% HPD intervals*

| | Dirichlet–Dirichlet | | Dirichlet–lognormal | | Uniform | |
|---|---|---|---|---|---|---|
| | **Mean** | **95% HPD** | **Mean** | **95%HPD** | **Mean** | **95% HPD** |
| $m_1$ | 0.097 | (0,0.280) | 0.101 | (0,0.282) | 0.099 | (0,0.250) |
| $m_2$ | 0.297 | (0.086,0.526) | 0.305 | (0.085,0.542) | 0.243 | (0.046,0.436) |
| $m_3$ | 0.3 | (0,0.514) | 0.324 | (0.045,0.586) | 0.258 | (0.035,0.475) |
| $m_4$ | 0.113 | (0,0.300) | 0.11 | (0,0.305) | 0.104 | (0,0.256) |
| $m_5$ | 0.061 | (0,0.196) | 0.07 | (0,0.195) | 0.081 | (0,0.198) |
| $m_6$ | 0.052 | (0,0.214) | 0.037 | (0,0.179) | 0.092 | (0,0.234) |
| $m_7$ | 0.079 | (0,0.270) | 0.053 | (0,0.230) | 0.123 | (0,0.289) |
| $\alpha_1$ | −0.494 | (−1.808,0.668) | −1.03 | (−3.005,0.961) | 0 | (0,0) |
| $\alpha_2$ | 0.113 | (−0.864,1.084) | 0.182 | (−1.191,1.680) | 0 | (0,0) |
| $\omega$ | 0.609 | (0.105,1.000) | 0.613 | (0.110,1.000) | 0.616 | (0.056,0.986) |
| $\rho$ | 0.118 | (0,0.343) | | | | |
| $\tau$ | | | 1.453 | (0.183,4.377) | | |

and posterior analysis indicates distance has a moderate effect on the proportion contribution.

The 95% HPD intervals of all regression coefficients in all models include zero, which is not surprising from the analysis of the simulation results. The large varia-tion in $\boldsymbol{\alpha}$ is presumably due to the small number of source populations. Nonetheless, there is some support for the notion that the coefficient associated with distance, $\alpha_1$, is negative. The posterior probability that $\alpha_1$ is negative is more than 0.785 for the Dirichlet–Dirichlet model and 0.850 for the Dirichlet–lognormal model. The posterior probability that $\alpha_2$ is positive in our model is 0.593 and 0.596 in the Dirichlet–lognormal model. The Dirichlet–Dirichlet model shows a shorter HPD interval compared to the Dirichlet–lognormal model, which is also consistent with the simulation results. In short, distance has a negative effect on the proportional contributions and population sizes have minor positive effects on the proportional contributions.

As discussed above, the parameter $\rho$ in our hierarchical model is analogous to a 'goodness of fit' measure for the relationship between the covariates and $\mathbf{m}$. Specifically, $1 - \rho$ is roughly the proportion of variance in $\mathbf{m}$ explained by the regression. As the results in Table 4 show, the posterior mean for $1 - \rho$ is near 0.9, which indicates a fairly tight regression in spite of the uncertainties associated with $\alpha$. In summary, we conclude that there is moderate support for the hypothesis that increasing distances between the source and colony populations decrease the proportional contributions of the sources to the colony.

## 5. Conclusions

The primary goal of this analysis is to incorporate environmental/demographic information into the estimation of the proportional contributions of source populations to a new colony through appropriate informative priors. Two other models are available which satisfy the constraint that the sum of the proportional contributions must equal one, i.e., additive logistic transformation [9] and Dirichlet–lognormal model [3]. We introduce a parametrization for the Dirichlet prior derived from population genetics in which we specify the mean, $\varphi_i$, and variance, $\rho(1 - \varphi_i)\varphi_i$, of the mixture parameters and a linear model for the parameters of a second Dirichlet that determines $\varphi_i$. The Dirichlet–Dirichlet prior has several advantages over the alternatives. First, the parameter $\rho$ has a natural vague prior distribution, a uniform distribution [0,1]. Second, $\rho$ controls the variance of the Dirichlet prior and $1 - \rho$ has a natural interpretation as the proportion of variance explained by regression. Finally, the mean of the proportional contributions is not affected by the parameter $\rho$ and the regression coefficients have a direct interpretation as regression effects on proportional representation. The separation of mean effect and variance effect is a major advantage of the proposed formulation compared to alternative models where the proportional contribution for any given population depends on the relative magnitude of coefficients associated with other regression components and their random effects.

The simulation study indicates that larger population divergence would lead to more precise estimation of the proportional contributions $\mathbf{m}$. Given a particular level of population divergence, better estimates of $\mathbf{m}$ can also be achieved by including more loci in the analysis. The simulations show that the Dirichlet–Dirichlet prior has better performance in estimating the regression coefficients in term of posterior variation than a Dirichlet–lognormal prior.

When we apply our model to the grey seal data we find that the distance between a source island and the new colony play a moderate role in its proportional contribution but that the effect of source population productivity is weak. These results are consistent with those presented in [3], but the posterior variability of the regression coefficients is smaller, as in the simulation study. The advantages of the formulation presented here seem likely to be generally available in the analysis of compositional data. In particular, a formulation similar to the one used here may be generally useful in modeling situations where additive logistic transformations have been the norm, both because of direct interpretability of regression coefficients and the natural interpretation of $1 - \rho$ as a goodness of fit measure.

## Appendix: A general approach for updating a proportional vector

The conditional distributions of model parameters are non-standard distributions; hence, we use a Metropolis–Hastings algorithm nested within Gibbs sampling to conduct each MCMC update. Several vector parameters, $\mathbf{P}$, $\mathbf{m}$, and $\varphi$, are subject to the constraint that the support of their components is on $[0,1]$ and the summation equals to one. We use a multidimensional logit transformation to 'de-constrain' the parameters and perform Metropolis–Hastings updating using a Normal proposal density. Let $\boldsymbol{\theta}$ be a vector of dimension $p+1$ with constraints $\theta_i > 0$ and $\sum_{i=1}^{p+1} \theta_i = 1$. Let

$$\theta_i = \frac{\exp(\xi_i)}{1 + \sum_{j=1}^{p} \exp(\xi_j)}.$$

The Jacobian matrix $\partial f(\boldsymbol{\theta})/\partial \boldsymbol{\xi}$ is the matrix with entries

$$x_{ij} = \begin{cases} \frac{e^{\xi_i} + e^{\xi_i}(\sum_{j=1}^{p} e^{\xi_j}) - e^{2\xi_i}}{(1 + \sum_{j=1}^{p} e^{\xi_j})^2}, & \text{for } i = j, \\ \frac{-e^{\xi_i + \xi_j}}{(1 + \sum_{j=1}^{p} e^{\xi_j})^2}, & \text{for } i \neq j. \end{cases}$$

It can be shown that the determinant of the Jacobian matrix is

$$\frac{e^{\sum_{j=1}^{p} \xi_j}}{(1 + \sum_{j=1}^{p} e^{\xi_j})^{p+1}}.$$

The full conditional distribution of $\xi$ is

$$f(\xi|D) = f(\boldsymbol{\theta}|D) \frac{e^{\sum_{j=1}^{p} \xi_j}}{(1 + \sum_{j=1}^{p} e^{\xi_j})^{p+1}}.$$

Instead of sampling $\boldsymbol{\theta}$, we conduct a Metropolis-Hastings update for $\boldsymbol{\xi}$ using a normal proposal density $N(\hat{\xi}, \hat{\sigma}_{\xi}^2)$, where $\hat{\xi}$ is the maximizer of $\pi(\xi|D)$ and $\hat{\sigma}_{\xi}^2$ is the estimated variance, which could be a fixed value based on a pilot run or the inverse of the score matrix. Alternatively, we can use a Normal proposal density centered at the current value. The algorithm operates as follows:

Step 1. Let $\boldsymbol{\xi}$ be the current value. Find the maximum likelihood estimate of $\boldsymbol{\xi}$, $\hat{\boldsymbol{\xi}}$.

Step 2. Generate a proposal value $\boldsymbol{\xi}^*$ from $N(\hat{\boldsymbol{\xi}}, \hat{\sigma}_{\boldsymbol{\xi}}^2)$.

Step 3. A move from $\boldsymbol{\xi}$ to $\boldsymbol{\xi}^*$ is made with probability

$$\min \left\{ \frac{f(\boldsymbol{\xi}^*|\mathbf{D}) \Phi(\frac{\boldsymbol{\xi} - \hat{\boldsymbol{\xi}}}{\sigma \boldsymbol{\xi}})}{f(\boldsymbol{\xi}|\mathbf{D}) \Phi(\frac{\boldsymbol{\xi}^* - \hat{\boldsymbol{\xi}}}{\sigma \boldsymbol{\xi}})}, 1 \right\}$$

where $\Phi$ is the standard normal probability density function. The $\xi$ is then converted back to its expression in terms of $\theta$.

**Acknowledgments.** We are indebted to William Amos for sharing the data on grey seals that motivated this study. Collection of these data was supported by NERC grant GR3/11662 to W. Amos.